\documentstyle[12pt]{article}
\makeatletter%  allow access to internal LaTeX commands

\renewcommand{\section}{\@startsection{section}{1}{\z@}%
    {-3.5ex plus -1ex minus -.2ex}%
    {2.3ex plus.2ex}%
    {\centering\normalsize\bf}}

\renewcommand{\subsection}{\@startsection{subsection}{2}{\z@}%
	{-3.25ex plus -1ex minus -.2ex}%
	{1.5ex plus .2ex}%
	{\centering\normalsize\it}}

\makeatother%  turn off access to internal LaTeX commands

\newenvironment{references}{%
	\newpage%  start new page
	\begin{thebibliography}{99}%
	\frenchspacing}%  use tighter spacing
	{\end{thebibliography}}

\makeatletter%  allow access to internal LaTeX commands
	\@addtoreset{equation}{section}%
\makeatother%  turn off access to internal LaTeX commands

\newcommand{\beq}{\begin{eqnarray}}% can be used as {equation} or {eqnarray}
\newcommand{\eeq}{\end{eqnarray}}

\def\endignore{}
\def\ignore #1\endignore{}

\newcommand{\mybar}[1]%
	{\kern 0.8pt\overline{\kern -0.8pt#1\kern -0.8pt}\kern 0.8pt}
\newcommand{\sla}[1]%
	{\raise.15ex\hbox{$/$}\kern-.57em #1}% Feynman slash
\newcommand{\roughly}[1]%
	{\mathrel{\raise.3ex\hbox{$#1$\kern-.75em\loIr1ex\hbox{$\sim$}}}}

\newcommand{\drawsquare}[2]{\hbox{%
\rule{#2pt}{#1pt}\hskip-#2pt%  left vertical
\rule{#1pt}{#2pt}\hskip-#1pt%  loIr horizontal
\rule[#1pt]{#1pt}{#2pt}}\rule[#1pt]{#2pt}{#2pt}\hskip-#2pt%  upper horizontal
\rule{#2pt}{#1pt}}% right vertical

\newcommand{\Yfund}{\raisebox{-.5pt}{\drawsquare{6.5}{0.4}}}%  fund
\newcommand{\Yasymm}{\raisebox{-3.5pt}{\drawsquare{6.5}{0.4}}\hskip-6.9pt%
        \raisebox{3pt}{\drawsquare{6.5}{0.4}}}%  antisymmetric second rank

\newcommand{\nop}[1]{:\kern-.3em#1\kern-.3em:}% normal ordered product

\newcommand{\Group}[2]{{\hbox{{\sl #1}($#2$)}}}
\newcommand{\U}[1]{\Group{U\kern0.05em}{#1}}
\newcommand{\SU}[1]{\Group{SU\kern0.1em}{#1}}
\newcommand{\SL}[1]{\Group{SL\kern0.05em}{#1}}
\newcommand{\Sp}[1]{\Group{Sp\kern0.05em}{#1}}
\newcommand{\SO}[1]{\Group{SO\kern0.1em}{#1}}

\newcommand{\jref}[4]{{\it #1} {\bf #2} (#4) #3}

\newcommand{\NPB}[3]{\jref{Nucl.\ Phys.}{B#1}{#2}{#3}}

\newcommand{\PLB}[3]{\jref{Phys.\ Lett.}{#1B}{#2}{#3}}

\newcommand{\PRD}[3]{\jref{Phys.\ Rev.}{D#1}{#2}{#3}}

\newcommand{\PRL}[3]{\jref{Phys.\ Rev.\ Lett.}{#1}{#2}{#3}}

\newcommand{\PTP}[3]{\jref{Prog.\ Theor.\ Phys.}{#1}{#2}{#3}}

\def\vbr{\vphantom{\sqrt{F_e^i}}}% vertical brace for tables

\newcommand{\spot}{superpotential}

\begin{document}

\pagestyle{empty}

\begin{titlepage}
\def\thepage {}        % Kill page numbering

\title{Duals for \SU{N} SUSY gauge theories\\
with an antisymmetric tensor: \\
five easy flavors}

\author{John Terning\thanks{e-mail: terning@alvin.lbl.gov}\\
\\
\baselineskip=10pt
{\normalsize \it Department of Physics}\\
{\normalsize \it University of California}\\
{\normalsize \it Berkeley, CA 94720}}

\baselineskip=12pt

\date{December 16, 1997}

\maketitle

\vspace*{-110mm}
\makebox[13cm][r]{UCB-PTH-97-31}\\
\makebox[13cm][r]{LBNL-41105}
\vspace*{110mm}

\abstract{
I consider ${\cal N}=1$ supersymmetric \SU{N_c} gauge theories with matter
fields consisting of one antisymmetric representation, five flavors, and
enough anti-fundamental representations to cancel the gauge anomaly.
Previous analyses are extended to the case of even $N_c$ with no
superpotential. Using holomorphy I show that the theory has an
interacting infrared fixed point for sufficiently large $N_c$.
These theories are interesting due to the fact that in going from five to four
flavors the theory goes from a non-trivial infrared fixed point to
confinement, in contradistinction to SUSY QCD, but in analogy to the behavior
expected in non-SUSY QCD.}

\vfill
\end{titlepage}

\baselineskip=18pt
\pagestyle{plain}
\setcounter{page}{1}

% ----------------------------------------------------------------------------
\section{Introduction}
% ----------------------------------------------------------------------------
In recent years our understanding of the infrared behavior of vector-like
${\cal N}=1$ supersymmetric (SUSY) gauge
theories has increased dramatically, primarily due to the work of Seiberg
\cite{Seib,IntSeib}.
In particular it is now known for SUSY QCD with a given number of flavors
whether the theory has: an unstable vacuum, a confined description, a weakly
coupled (infrared free) dual gauge description, a non-trivial infrared fixed
point, or a trivial infrared fixed point. Some work has also been done on
chiral SUSY gauge theories, but our understanding of these more complex
theories
is far from complete.
Chiral SUSY gauge theories are of special interest since they can dynamically
break SUSY, unlike most theories with vector matter\footnote{For a vector-like
theory that dynamically breaks SUSY see ref. \cite{IYIT}.}. Among the simplest
chiral theories are
those with an antisymmetric tensor. Consider \SU{N_c} with one
antisymmetric tensor, $(N_c-4)$  $\overline{\bf N_c}$'s and $F$  flavors (a
flavor is one ${\bf N_c}$ and one $\overline{\bf N_c}$);  it is known that
this
theory is confining \cite{Poppitz,Pouliot,sconf} for $F=3$ or 4.  Thus the
simplest example of this type of chiral SUSY theory which admits a dual gauge
description is $F=5$.  What is unknown is whether the theory has an infrared
free
dual gauge description or an interacting infrared fixed point.  In this paper
I show how to use the ``deconfinement'' method introduced by Berkooz
in ref. \cite{Berkooz} and elaborated in refs. \cite{Pouliot,decon,LST} to
construct simple duals for the case $F=5$  (pointing out why this case is
special) and $N_c$ even and compare with the previously know dual for  odd
$N_c$.    Using holomorphy I show that the dual
(and hence the original theory) has an interacting infrared fixed point at the
origin of moduli space for sufficiently large $N_c$.
Finally I present my conclusions, and discuss the analogous behavior in
non-SUSY QCD.

% ----------------------------------------------------------------------------
\section{Duality for \SU{2N}}
% ----------------------------------------------------------------------------
The theory I wish to study has gauge group \SU{N_c} with $5$ chiral
superfields  $q$ in the (defining) ${\bf N_{c}}$ representation, one matter
field $A$ in the antisymmetric tensor representation, and $N_c+1$ matter fields
$\overline{ q}$ in the $\overline{\bf N_{c}}$ representation.
This theory has the anomaly-free global symmetry $\SU{5} \times \SU{N_c+1}
\times
\U{1}_R \times \U{1}_X \times \U{1}_Y$.
The field content (with global charges) is given in Table 1.

% TABLE 1
\begin{table}[htbp]
\centering
\begin{tabular}{c|c|ccccc}
 & \SU{N_c} & \SU{5}  & \SU{N_c+1} & $\U{1}_R$ & $\U{1}_X$ & $U(1)_Y$
\\
\hline
$q$ & \Yfund & \Yfund & {\bf 1}  & ${{4}\over{N_c+6}}$  & $1$ & $-1\vbr$ \\
$\bar{ q}$ &  $\overline{\Yfund}$  & {\bf 1}  & \Yfund & ${{4}\over{N_c+6}}$  &
$-1$ & $-1\vbr$ \\
$A$ & \Yasymm & {\bf 1} & {\bf 1} & $0$  & ${{N_c-4}\over{N_c-2}}\vbr$ &
${{N_c+6}\over{N_c-2}}\vbr$ \\
\end{tabular}
\label{Sp}
\parbox{4in}{\caption{Field content of the theory.}}
\end{table}

This theory has been considered previously by Berkooz \cite{Berkooz} for
even $N_c (=2N)$, with the addition of a superpotential $W= {\rm
Pf}(A)$. I will consider this theory with no \spot. The case $N_c$  odd with no
superpotential has been discussed by Pouliot \cite{Pouliot} (see also ref.
\cite{chou}).

I can replace the antisymmetric tensor  by a composite ``meson'' operator of a
confining \Sp{2N-2} group:
\beq
A^{ab} \to x^{aa'} x^{bb'}\,J_{a' b'},
\eeq
where $a, b$ are \SU{2N} indices and $a',b'$ are \Sp{2N-2} indices and $J_{a'
b'}$
is the invariant tensor.
I must also introduce additional fields that
transform under \Sp{2N-2} and add terms to the \spot\ in the deconfined
description.
The matter content of the model that accomplishes this is displayed in
Table 2.

% TABLE 2
\begin{table}[htbp]
\label{SUSp}
\centering
\begin{tabular}{c|cc|cccccc}
 & \SU{N_c} & \Sp{D} & $\SU{2}_f$ & \SU{5} & \SU{C} & $\U{1}_R$ &
$\U{1}_X$ & $U(1)_Y$
\\
\hline
$q$ & \Yfund & {\bf 1} & {\bf 1} & \Yfund &  {\bf 1} & ${{2}\over{N+3}}$ & $1$
& $-1\vbr$ \\
$\bar{q}$ &$ \overline{\Yfund}$ & {\bf 1} &  {\bf 1} & {\bf 1} & \Yfund
&${{2}\over{N+3}}$ & $-1$ & $-1\vbr$ \\
$x$ & \Yfund & \Yfund & {\bf 1}  & {\bf 1} & {\bf 1}  & $0\vbr$ &
${{N-2}\over{2N-2}}\vbr$ & ${{N+3}\over{2N-2}}\vbr$ \\
$p$ &$ \overline{\Yfund}$ & {\bf 1} & \Yfund &  {\bf 1} &  {\bf 1} & 1 &
${{N-2}\over{2}}\vbr$  & ${{N+3}\over{2}}\vbr$\\
$r$ & {\bf 1} & \Yfund & \Yfund & {\bf 1} &  {\bf 1} & 1 &
${{-N(N-2)}\over{2N-2}}\vbr$  & ${{-N(N+3)}\over{2N-2}}\vbr$ \\
$s$ & {\bf 1} & {\bf 1} & {\bf 1} & {\bf 1} &  {\bf 1} & 0 &
${{N(N-2)}\over{N-1}}\vbr$ & ${{N(N+3)}\over{N-1}}\vbr$ \\
\end{tabular}
\parbox{4in}{\caption{Field content of the ``deconfined" theory, where $D =
2N-2$, and $C = 2N+1$.}}
\end{table}
The \spot\ in the ``deconfined" description is
\beq
W = x  r  p + r r s.
\eeq
(I have set the coefficients of the superpotential to $1$ by rescaling
the fields.)
The purpose of the \spot\ is to remove the unwanted ``meson''
states $(x r)$ and $(r r)$ that appear when the \Sp{2N-2} group
confines. These two ``mesons" get masses with $p$ and $s$ respectively.
Note that, as discussed in ref.~\cite{LST}, gauge anomaly cancellation for
\Sp{2N -2}
forces the fields $r$ and $p$ to have a fictitious global $\SU{2}_f$ symmetry.
This symmetry is fictitious in the sense that
none of the physical low energy degrees of freedom transform under it: $r$ is
confined,
and $p$ is massive. This symmetry will be useful later in determining which of
several dual
descriptions might be useful.

I can now use the known dual description of \SU{2N} gauge theory with
fundamentals \cite{Seib} to write a dual description of this theory
in terms of a theory with gauge group $\SU{3} \times \Sp{2N -2}$.
The field content of this dual is given in Table 3.
% TABLE 3
\begin{table}[htp]
\label{SU3Sp}
\centering
\begin{tabular}{c|cc|cccccc}
 & \SU{3} & \Sp{D} & $\SU{2}_f$ & \SU{5} & \SU{C} & $\U{1}_R$ & $\U{1}_X$
& $U(1)_Y$
\\
\hline
$q_1$ & \Yfund & {\bf 1} & {\bf 1} & $\overline{\Yfund}$ &  {\bf 1} &
${{4}\over{3(N+3)}}$ & ${{N}\over{3}}$ & ${{N+1}\over{3}}\vbr$ \\
$\bar{q_1}$ &$ \overline{\Yfund}$ & {\bf 1} &  {\bf 1} & {\bf 1} & $\overline{
\Yfund} $&${{6N+2}\over{3(N+3)}}$ & $-{{N}\over{3}}$ & ${{5-N}\over{3}}\vbr$ \\
$x_1$ & \Yfund & \Yfund & {\bf 1}  & {\bf 1} & {\bf 1}  &
${{10}\over{3(N+3)}}\vbr$ & ${{N(2N+1)}\over{3(2N-2)}}\vbr$ &
${{(2N+1)(N-5)}\over{3(2N-2)}}\vbr$ \\
$p_1$ &$ \overline{\Yfund}$ & {\bf 1} & \Yfund &  {\bf 1} & {\bf 1}
&${{3N-1}\over{3(N+3)}}\vbr$  &${{-5N(N-1)}\over{3(2N-2)}}$ &
${{-5(N+1}\over{6}}\vbr$ \\
$r$ & {\bf 1} & \Yfund & \Yfund & {\bf 1} &  {\bf 1} & $1\vbr$
&${{-N(N-2)}\over{2N-2}}$ &${{-N(N+3)}\over{2N-2}}\vbr$ \\
$s$ & {\bf 1} & {\bf 1} & {\bf 1} & {\bf 1} &  {\bf 1} & $0\vbr$
&${{N(N-2)}\over{N-1}}$ &${{N(N+3)}\over{N-1}}\vbr$ \\
$(\bar{q}q)$ & {\bf 1} & {\bf 1} & {\bf 1} &  \Yfund &  \Yfund &
${{4}\over{N+3}}\vbr$ & 0 & -2 \\
$(xp)$ & {\bf 1} & \Yfund &  \Yfund & {\bf 1} &  {\bf 1} & $1\vbr$
&${{N(N-2)}\over{2N-2}}$ &${{N(N+3)}\over{2N-2}}\vbr$ \\
$(qp)$ & {\bf 1} & {\bf 1} & \Yfund & \Yfund &  {\bf 1}
&${{N+5}\over{N+3}}\vbr$ &${{N}\over{2}}$ & ${{N+1}\over{2}}$ \\
$(\bar{q}x)$ & {\bf 1} & \Yfund & {\bf 1} & {\bf 1} &  \Yfund &
${{2}\over{N+3}}\vbr$ &${{-N}\over{2N-2}}$ &${{5-N}\over{2N-2}}$ \\
\end{tabular}
\parbox{4in}{\caption{Field content of the first dual description, where
$D=2N-2$, $C=2N+1$.}}
\end{table}
This dual has a \spot
\beq
W & = & (xp)r +rrs + (\bar{q}q) q_1 \bar{ q_1} +
(xp)x_1p_1+(qp)q_1p_1+(\bar{q}x) \bar{ q_1}x_1~.
\eeq
I have introduced some notation here to simplify the later exposition: $q_i$
refers to the field which is the ``dual" of $q_{i-1}$, where $q_0 \equiv q$,
and I
will denote
a  ``meson" which is the mapping of $q_{i} p_{j}$ (and couples to $q_{i+1}$ and
$p_{j+1} $ in the dual \spot) by $(q_i p_j)$.  For later convenience I will
relabel the ``meson" $(\overline{q} x)$ by  $y$.
The massive fields $(x p)$ and $r$ can be integrated out, leaving the
superpotential:
\beq
\label{firstdualW}
W & = & x_1^2 p_1^2 s + (\bar{q}q) q_1 \bar{ q_1} +(qp)q_1p_1+y \bar{ q_1}x_1~.
\eeq
The anomaly matching is guaranteed to work by the anomaly matching of the
{\sl SU} duality used in its construction.
%\footnote{The anomaly matching is explicitly checked in the Appendix.}.

The dual description obtained above is, unfortunately, almost completely
useless, since it possesses a fictitious global $\SU{2}_f$ symmetry: any field
which transforms under this fictitious symmetry must either be massive or
strongly coupled since it cannot appear in the physical spectrum. It is obvious
that  additional dual descriptions can be obtained by alternating the gauge
group that  duality is applied to \cite{Berkooz,Pouliot,LST}; what is
surprising is that such an exercise turns out to be useful.  Going through a
repeated application of alternating dualities produces (for the case of five
flavors) a dual with no
fields transforming under $\SU{2}_f$.  The remainder of this section is
devoted to detailing this procedure, the reader who is
interested in results rather than techniques is urged to skip  ahead
to Table 6 where the final dual is presented.

The next step is to apply duality to the \Sp{2N-2} gauge
group. The field content of the resulting dual is given in Table 4,
and the
superpotential is
\beq
W & = & z p_1^2 s + (\bar{q}q) q_1 \bar{ q_1} +(qp)q_1p_1+(y x_1)\bar{ q_1} \\
\nonumber
& &+z x_2 x_2 + (yy)y_1 y_1 +(x_1 y)y_1 x_2~,
\eeq
where I have renamed $(x_1 x_1)$ to be $z$.
% TABLE 4
\begin{table}[htbp]
\label{SU3SU2}
\centering
\begin{tabular}{c|cc|cccccc}
 & \SU{3} & \SU{2} & $\SU{2}_f$ & \SU{5} & \SU{C} & $\U{1}_R$ & $\U{1}_X$
& $U(1)_Y$
\\
\hline
$q_1$ & \Yfund & {\bf 1} & {\bf 1} & $\overline{ \Yfund}$ &  {\bf 1} &
${{4}\over{3(N+3)}}$ & ${{N}\over{3}}$ & ${{N+1}\over{3}}\vbr$ \\
$\bar{q_1}$ &$ \overline{\Yfund}$ & {\bf 1} &  {\bf 1} & {\bf 1} & $ \overline{
\Yfund} $ &${{6N+2}\over{3(N+3)}}$ & $-{{N}\over{3}}$ & ${{5-N}\over{3}}\vbr$
\\
$x_2$ & $\overline{ \Yfund}$ & \Yfund & {\bf 1}  & {\bf 1} & {\bf 1}  &
${{3N-1}\over{3(N+3)}}\vbr$ & ${{-N(2N+1)}\over{3(2N-2)}}\vbr$ &
${{-(2N+1)(N-5)}\over{3(2N-2)}}\vbr$ \\
$p_1$ &$ \overline{\Yfund}$ & {\bf 1} & \Yfund &  {\bf 1} & {\bf 1}
&${{3N-1}\over{3(N+3)}}\vbr$  &${{-5N(N-1)}\over{3(2N-2)}}$ &
${{-5(N+1)}\over{6}}\vbr$ \\
$s$ & {\bf 1} & {\bf 1} & {\bf 1} & {\bf 1} &  {\bf 1} & $0\vbr$
&${{N(N-2)}\over{N-1}}$ &${{N(N+3)}\over{N-1}}\vbr$ \\
$(\bar{q}q)$ & {\bf 1} & {\bf 1} & {\bf 1} &  \Yfund &  \Yfund &
${{4}\over{N+3}}\vbr$ & 0 & -2 \\
$(qp)$ & {\bf 1} & {\bf 1} & \Yfund & \Yfund &  {\bf 1}
&${{N+5}\over{N+3}}\vbr$ &${{N}\over{2}}$ & ${{N+1}\over{2}}$ \\
$y_1$ & {\bf 1} & \Yfund & {\bf 1} & {\bf 1} & $ \overline{ \Yfund} $ &
${{N+1}\over{N+3}}\vbr$ &${{N}\over{2N-2}}$ &${{N-5}\over{2N-2}}$ \\
$z$ & $\overline{ \Yfund}$ & {\bf 1} & {\bf 1} & {\bf 1} &  {\bf 1}  &
${{20}\over{3(N+3)}}\vbr$ & ${{N(2N+1)}\over{3(N-1)}}\vbr$ &
${{(2N+1)(N-5)}\over{3(N-1)}}\vbr$ \\
$(yy)$ & {\bf 1} & {\bf 1} & {\bf 1} & {\bf 1} &  \Yasymm &
${{4}\over{N+3}}\vbr$ &${{-N}\over{N-1}}$ &${{5-N}\over{N-1}}$ \\
$(x_1 y)$ & \Yfund & {\bf 1} & {\bf 1} & {\bf 1} &  \Yfund &
${{16}\over{3(N+3)}}\vbr$ &${{N}\over{3}}$ &${{N-5}\over{3}}\vbr$ \\
\end{tabular}
\parbox{4in}{\caption{Field content of the second dual description, where
$C=2N+1$.}}
\end{table}

After integrating out $\overline{q_1}$ and $(x_1 y)$, the superpotential is:
\beq
W & = & z p_1^2 s - (\bar{q}q) q_1 y_1 x_2 +(qp)q_1p_1
+z x_2 x_2 + (yy)y_1 y_1 ~.
\eeq
At this point it can be seen why the case of 5 flavors is so special. If the
analysis so far had been done for $F$ flavors, then the gauge group \SU{3}
would instead be \SU{F-2}, and the field
$z$ would be an antisymmetric tensor\footnote{The antisymmetric tensor for
\SU{3} is just a $\bar{\bf 3}$.}.  Then to further dualize the \SU{F-2} would
require the introduction of an additional
``deconfinement" module, and hence an even more complicated description of the
theory.

I can now use the known dual of an \SU{3} gauge theory with fundamental
representations
to find another dual, with the field
content given in
Table 5; the superpotential is
\beq
W & = & x_3^2 s - (\bar{q}q) (q_1  x_2) y_1 +(qp)(q_1p_1 )
+p_2^2 + (yy)y_1 y_1 +(q_1 x_2)q_2 x_3 \\ \nonumber
& & +(q_1 p_1) q_2 p_2 +(q_1 z) q_2 z_1~.
\eeq
% TABLE 5
\begin{table}[htbp]
\label{SU2SU2}
\centering
\begin{tabular}{c|cc|cccccc}
 & \SU{2} & \SU{2} & $\SU{2}_f$ & \SU{5} & \SU{C} & $\U{1}_R$ & $\U{1}_X$
& $U(1)_Y$
\\
\hline
$q_2$ & \Yfund & {\bf 1} & {\bf 1} &  \Yfund &  {\bf 1} & ${{2}\over{N+3}}$ &
${{N}\over{2}}\vbr$ & ${{N+1}\over{2}}\vbr$ \\
$x_3$ &  \Yfund  & \Yfund & {\bf 1}  & {\bf 1} & {\bf 1}  & $1\vbr$ &
${{-N(N-2)}\over{2N-2}}$ & ${{-N(N+3)}\over{2N-2}}\vbr$ \\
$p_2$ & \Yfund  & {\bf 1} & \Yfund &  {\bf 1} & {\bf 1} & $1\vbr$  & $0$& $0$\\
$s$ & {\bf 1} & {\bf 1} & {\bf 1} & {\bf 1} &  {\bf 1} & $0\vbr$
&${{N(N-2)}\over{N-1}}$ &${{N(N+3)}\over{N-1}}\vbr$ \\
$(\bar{q}q)$ & {\bf 1} & {\bf 1} & {\bf 1} &  \Yfund &  \Yfund &
${{4}\over{N+3}}\vbr$ & 0 & -2 \\
$(qp)$ & {\bf 1} & {\bf 1} & \Yfund & \Yfund &  {\bf 1}
&${{N+5}\over{N+3}}\vbr$ &${{N}\over{2}}$ & ${{N+1}\over{2}}$ \\
$y_1$ & {\bf 1} & \Yfund & {\bf 1} & {\bf 1} & $ \overline{ \Yfund} $ &
${{N+1}\over{N+3}}\vbr$ &${{N}\over{2N-2}}$ &${{N-5}\over{2N-2}}$ \\
$z_1$ &  \Yfund & {\bf 1} & {\bf 1} & {\bf 1} &  {\bf 1} &
${{2N-4}\over{N+3}}\vbr$ &${{-N(3N-1)}\over{2N-2}}$ &${{5+6N-3N^2}\over{2N-2}}$
\\
$(yy)$ & {\bf 1} & {\bf 1} & {\bf 1} & {\bf 1} &  \Yasymm
&${{4}\over{N+3}}\vbr$ &${{-N}\over{N-1}}$ &${{5-N}\over{N-1}}$ \\
$(q_1 x_2)$ & {\bf 1} &  \Yfund & {\bf 1} &$\overline{ \Yfund}$ & {\bf 1}
&${{N+1}\over{N+3}}\vbr$ &${{-N}\over{2N-2}}$ &${{3N+1}\over{2N-2}}$ \\
$(q_1 p_1)$ & {\bf 1} & {\bf 1} & \Yfund &$\overline{ \Yfund}$& {\bf 1}
&${{N+1}\over{N+3}}\vbr$ &${{-N}\over{2}}$ &${{-N-1}\over{2}}$ \\
$(q_1 z)$ & {\bf 1} & {\bf 1} & {\bf 1} & $\overline{ \Yfund}$ & {\bf 1}
&${{8}\over{N+3}}\vbr$ &${{N^2}\over{N-1}}$ &${{N^2-3N-2}\over{N-1}}$ \\
\end{tabular}
\parbox{4in}{\caption{Field content of the third dual description, where
$C=2N+1$.}}
\end{table}

Note the ``baryonic" operator mapping:
\beq
\begin{array}{clc}
z p_1 p_1 & \to & x_3^2,\\
z x_2 x_2 & \to & p_2^2,\\
x_2 p_1 p_1 & \to & z_1x_3,\\
x_2 x_2 p_1 & \to & z_1p_2~.
\end{array}
\eeq
After integrating out $p_2$, $(qp)$, and $(q_1 p_1)$ there are no longer any
fields that transform under the fictitious global $\SU{2}_f$ symmetry, although
the singlet field $s$ still remains.  The superpotential is given by

\beq
W & = & x_3^2 s - (\bar{q}q) (q_1  x_2) y_1 + (yy)y_1 y_1 +(q_1 x_2)q_2 x_3 \\
\nonumber
& & +(q_1 p_1) q_2 p_2 +(q_1 z) q_2 z_1~.
\eeq

To obtain a slightly simpler dual description, I can now apply duality one more
time to the first \SU{2} gauge group.
After integrating out a number of massive fields I find the field content given
in
Table 6 with a superpotential given by:
\beq
W & = (\bar{q}q) q_3  x_4 y_1  + (yy)y_1 y_1 +(q_2 q_2)q_3 q_3 +(x_3 z_1)
x_4z_2 ~.
\label{finalspot}
\eeq

% TABLE 6
\begin{table}[htbp]
\label{SU2SU2again}
\centering
\begin{tabular}{c|cc|ccccc}
 & $\SU{2}_1$ & $\SU{2}_2$  & \SU{5} & \SU{2N+1} & $\U{1}_R$ & $\U{1}_X$ &
$U(1)_Y$
\\
\hline
$q_3$ & \Yfund & {\bf 1} &$\overline{  \Yfund} $ &  {\bf 1} &
${{N+1}\over{N+3}}$ & ${{-N}\over{2}}$ & ${{-N-1}\over{2}}\vbr$ \\
$x_4$ &  \Yfund  & \Yfund   & {\bf 1} & {\bf 1}  & $0\vbr$ &
${{N(N-2)}\over{2N-2}}$ & ${{N(N+3)}\over{2N-2}}\vbr$ \\
$y_1$ & {\bf 1} & \Yfund  & {\bf 1} & $ \overline{ \Yfund} $  &
${{N+1}\over{N+3}}\vbr$ &${{N}\over{2N-2}}$ &${{N-5}\over{2N-2}}$  \\
$z_2$ &  \Yfund & {\bf 1}  & {\bf 1} &  {\bf 1} & ${{7-N}\over{N+3}}\vbr$
&${{N(3N-1)}\over{2N-2}}$ & ${{3N^2-6N-5}\over{2N-2}}$ \\
$(x_3 z_1)$ & {\bf 1} & \Yfund & {\bf 1}  &  {\bf 1} & ${{3N-1}\over{N+3}}\vbr$
&${{-N(4N-3)}\over{2N-2}}$ & ${{5+3N-4N^2}\over{2N-2}}$ \\
$(\bar{q}q)$ & {\bf 1} & {\bf 1}  &  \Yfund &  \Yfund & ${{4}\over{N+3}}\vbr$ &
0 & -2 \\
$(q_2 q_2)$ & {\bf 1} & {\bf 1} & \Yasymm & {\bf 1}    & ${{4}\over{N+3}} \vbr$
& $N$ & $N+1$ \\
$(yy)$ & {\bf 1} & {\bf 1}  & {\bf 1} &  \Yasymm &${{4}\over{N+3}}\vbr$
&${{-N}\over{N-1}}$ &${{5-N}\over{N-1}}$ \\
\end{tabular}
\parbox{4in}{\caption{Field content of the ``final" dual description.}}
\end{table}

The operator mapping of the chiral ring  is:
\beq
\begin{array}{clc}
\overline{q}q & \to & (\overline{q}q) ,\\
\overline{q} A \overline{q} & \to & (yy),\\
q A^{N-1} q & \to & (q_2 q_2),\\
q^4 A^{N-2}  & \to & q_3 z_2,\\
A^N & \to & x_4^2,\\
\overline{q}^{2N} & \to & y_1 (x_3 z_1) ~,
\end{array}
\eeq
where $y \equiv (\overline{q} x)$ and $z \equiv (x_1 x_1)$.

This dual has a simple relation to the spectrum of the confined description
found by Pouliot \cite{Pouliot} for the case of four flavors. Adding a mass
term for one flavor in the
original theory gives  breaks the flavor symmetry to \SU{4} $\times$ \SU{2N}.
In the dual description the mass term maps to a linear term for
$(\overline{q}q)$, which induces a vev for for the product $q_3 x_4 y_1$.
$D$-flatness ensures that, in an appropriate basis, each of these three fields
has only one non-zero component.  These vevs break the gauge symmetries
completely and produce mass terms for extra components of $(\overline{q}q)$,
$(yy)$,
and $(q_2 q_2)$ with uneaten pieces of $q_3$ and $y_1$. The vev of $x_4$ gives
a mass to one component each of $(x_3 z_1)$ and $z_2$, leaving two massless
fields, and one component of $x_4$ remains uneaten.
The operator mapping for four flavors is:
\beq
\begin{array}{clc}
\overline{q}q & \to & (\overline{q}q) ,\\
\overline{q} A \overline{q} & \to & (yy),\\
q A^{N-1} q & \to & (q_2 q_2),\\
q^4 A^{N-2}  & \to & \widehat{z_2},\\
A^N & \to & \widehat{x_4},\\
\overline{q}^{2N} & \to & \widehat{(x_3 z_1)} ~,
\end{array}
\eeq
where the $\, \widehat{ \, }\,$ superscript indicates the remaining massless
(singlet) component.
Poppitz and Trivedi \cite{Poppitz} showed that these theories can break SUSY
with the addition of some singlet fields, some superpotential terms, and the
gauging of a chiral $U(1)$ symmetry.  With corresponding manipulations of
the dual, it can  provide a weakly coupled description of their SUSY breaking
models.
% ----------------------------------------------------------------------------
\section{Comparison with Duality for \SU{2M-1}}
% ----------------------------------------------------------------------------

The case of odd $N_c$ has been studied previously by Pouliot \cite{Pouliot}.
In this section I will briefly review his dual in order make comparisons with
the even $N_c$ case discussed above. The field content (with global charges) is
given in Table 1, with $N_c = 2M-1$.
Pouliot deconfined the antisymmetric tensor with \Sp{2M-4} by introducing
fields $x$, $r$, and $p$ (as discussed earlier for the even case) with a
superpotential $W = xrp$.  The odd $N_c$ case is much simpler than the even
case because no fictitious global symmetry is needed nor is a singlet field
required. Pouliot then dualized \SU{2M-1} to \SU{2} in the usual fashion, and
further dualized \Sp{2M-4}  to  \SU{2}. After integrating out massive fields,
he arrived at a dual  with a superpotential given by:
\beq
W & =  (\bar{q}q) q_1  y_1  x_2+ (yy)y_1 y_1 +(q p)q_1 p_1 +(x_1 x_1) x_2 x_2
{}~,
\eeq
and the field content shown in Table 7 (using the notation  $y \equiv
(\overline{q} x))$.

% TABLE 7
\begin{table}[htbp]
\label{SU2SU2dejavu}
\centering
\begin{tabular}{c|cc|ccccc}
 & \SU{2} & \SU{2}  & \SU{5} & \SU{2M} & $\U{1}_R$ & $\U{1}_X$ & $U(1)_Y$
\\
\hline
$q_1$ & \Yfund & {\bf 1} &$\overline{  \Yfund} $ &  {\bf 1} &
${{6}\over{2M+5}}$ & ${{(M-1)(2M-1)}\over{4M-6}}$ &
${{2M^2+-5M-1}\over{4M-6}}\vbr$ \\
$x_2$ &  \Yfund  & \Yfund   & {\bf 1} & {\bf 1}  & ${{2M-5}\over{2M+5}} \vbr$ &
${{-M(2M-1)}\over{4M-6}}$ & ${{-M(2M-11)}\over{4M-6}}\vbr$ \\
$y_1$ & {\bf 1} & \Yfund  & {\bf 1} & $ \overline{ \Yfund} $  &
${{2M+1}\over{2M+5}}\vbr$ &${{2M-1}\over{4M-6}}$ &${{2M-11}\over{4M-6}}$  \\
$p_1$ &  \Yfund & {\bf 1}  & {\bf 1} &  {\bf 1} & ${{4M}\over{2M+5}}\vbr$
&${{-6M^2+13M-5}\over{4M-6}}$ & ${{-6M^2+3M+5}\over{4M-6}}$ \\
$(x_1 x_1)$ & {\bf 1} & {\bf 1} & {\bf 1}  &  {\bf 1} & ${{20}\over{2M+5}}\vbr$
&${{2M(2M-1)}\over{4M-6}}$ & ${{2M(2M-11)}\over{4M-6}}$ \\
$ (\bar{q}q)$ & {\bf 1} & {\bf 1}  &  \Yfund &  \Yfund & ${{8}\over{2M+5}}\vbr$
& 0 & -2 \\
$(q p)$ & {\bf 1} & {\bf 1} & \Yfund & {\bf 1}    & ${{4}\over{2M+5}} \vbr$ &
${{4M^2-10M+4}\over{4M-6}}$ & ${{4M^2+2M-4}\over{4M-6}}$ \\
$(yy)$ & {\bf 1} & {\bf 1}  & {\bf 1} &  \Yasymm &${{8}\over{2M+5}}\vbr$
&${{2-4M}\over{4M-6}}$ &${{22-4M}\over{4M-6}}$ \\
\end{tabular}
\parbox{4in}{\caption{Field content of the dual description for the case
$N_c=2M-1$.}}
\end{table}

The operator mapping is:
\beq
\begin{array}{clclc}
\overline{q}q & \to &  (\overline{q}q) ,\\
\overline{q} A \overline{q} & \to & (yy),\\
q A^{M-1}  & \to & (q p),\\
q^3 A^{M-2}  & \to & q_1 q_1,\\
\overline{q}^{2M-1} & \to & y_1 x_2 p_1 ,\\
\end{array}
\eeq
and for $M \ge 3$
\beq
\begin{array}{clclc}
q^{5} A^{M-3} & \to & (x_1 x_1)~.
\end{array}
\eeq

Even though only three dualities were required in the derivation of the odd
$N_c$ case, as opposed to five dualities for  even $N_c$,  the resulting dual
descriptions are quite similar.  The five or six  flavor models are special
for odd $N_c$ as well, since for a larger number of flavors the dual contains
tensor representations.

% ----------------------------------------------------------------------------
%\section{Dynamical SUSY Breaking}
% ----------------------------------------------------------------------------
%Poppitz and Trivedi \cite{Poppitz} showed that the $N_c=2N$ theory with the
% addition of some %singlet fields, some superpotential terms, and the gauging
% of a chiral $U(1)$ symmetry %dynamically breaks SUSY.  The theory they
% discussed had one flavor, so it is necessary to %add mass terms for four
% flavors, leaving one $q$ and $2N-3$ $\bar{q}$'s.  The mass terms %break the
% global symmetries down to \SU{2N-3} and three $U(1)$'s.  Poppitz and
% Trivedi's %gauged $U(1)$ corresponds to a linear combination of the two
% non-$R$ $U(1)$ symmetries with %charges for ($q$,$\overline{q}$,$A$) taken to
% be ($1-2N$,$-1$,$2$).  In order to cancel the %$U(1)^3$ gauge anomaly it is
% necessary to add $2N-3$ singlets, $S^i$ with $U(1)$ charge %$2N$.  The moduli
% space is then parameterized by $\overline{q} q A^N$, $S \overline{q} q$ and
% %$\overline{q} A \overline{q}$.  They then added to the superpotential two
% Yukawa terms:
%\beq
%W_{\rm yuk} = \gamma^{ij} \bar{q^i} A \bar{q^j} + \lambda^{ij}S^i  \bar{q}^j
% q~.
%\eeq
%In the dual description the mapping of the mass and Yukawa terms is
%\beq
%W= m \sum_{a=2}^{5}(\bar{q}^{2N-4+a} q^{a}) +\gamma^{ij} (y^i y^j) +
% \lambda^{ij} S_i %(\bar{q}^j q^1)
%\eeq

% ----------------------------------------------------------------------------
\section{The Infrared Fixed Point}
% ----------------------------------------------------------------------------
I would now like to demonstrate that the dual of the \SU{2N} theory described
above
(and the original theory itself)
has a non-trivial infrared fixed point at the origin of moduli space.  The
situation
is more difficult than in SUSY QCD since there are two gauge groups in the
dual.
The analysis can be simplified by using the fact that the ratio of the two
intrinsic
scales, $\Lambda_1$  and $\Lambda_2$ (corresponding to the $\SU{2}_1$ and
$\SU{2}_2$ gauge groups in the final dual
description given in Table 6), can be varied arbitrarily.  Holomorphy
\cite{holo} requires that,
aside
from singular points,  there can be no phase transitions as this ratio is
varied.
%$\Lambda_2$ is related to $\Lambda$ by \cite{Seib,IntSeib}:
%\beq
%\Lambda_2^{4-N} \Lambda^{2N-1} = \mu^{N+3}~.
%\eeq
There are two cases\footnote{For $N=4$, the holomorphic gauge coupling does not
run,
and can be set to be arbitrarily small.}
to consider: for $N > 4$ the $\SU{2}_2$ gauge group is infrared
free and $\Lambda_1 \ll \Lambda_2$  corresponds to the $\SU{2}_2$ gauge
coupling $g$ (renormalized at a scale near $\Lambda_1$) becoming arbitrarily
small, for
$N<4$ the $\SU{2}_2$ gauge group is asymptotically free and the limit
$\Lambda_1 \gg \Lambda_2$ also corresponds to  weak coupling for $\SU{2}_2$.
In both cases the gauge coupling $g \rightarrow 0$
as $\Lambda_1 / \Lambda_2 \rightarrow$ 0 or
$\infty$. Of course  $g$ cannot be simply set to zero for at least two reasons.
Firstly, the massless
spectrum is discontinuous in this limit, since setting $g=0$ causes
$D$-terms to vanish, thus enlarging the moduli space. More importantly
non-perturbative effects from the $\SU{2}_1$ gauge interactions can affect the
running
of $g$ in the infrared.  Thus a careful study is required.

Before proceeding with the details of the calculation I will sketch an outline
of the analysis.
I will analyze the dual at a renormalization scale somewhat
below the interaction scale of  $\SU{2}_1$  (i.e. $\mu < \Lambda_1$)
with an arbitrarily small (but non-zero) value for $g(\mu)$.
%This can be arranged
%by taking $\Lambda_2 \gg \Lambda_1$ for $N>4$ and $\Lambda_2 \ll \Lambda_1$
% for $N<4$.
At this scale the theory can be studied with perturbation theory in $g(\mu)$,
and at lowest order in $g(\mu)$ I will show that the $\SU{2}_1$ has a
non-trivial infrared fixed point.
I will then proceed to show that for sufficiently large $N$ the $\SU{2}_2$
interactions
are infrared free at this scale, i.e. that coupling $g$ has a trivial infrared
stable fixed point.  This is sufficient to prove that the theory with an
arbitrary ratio $\Lambda_1 / \Lambda_2$ has the same infrared fixed point,
since there cannot be a phase transition in the space of holomorphic couplings
\cite{holo}.  Thus the infrared limit can be understood simply through a
perturbative
analysis in the coupling $g$.

It is instructive to first consider the theory at zero-th order in $g$ (i.e.
with $g$ set to zero).
With  $\SU{2}_2$ turned off, the fields $y_1$ and  $(x_3 z_1)$ as well as the
products $q_3 x_4$  and $x_4 z_2$ become gauge
invariant operators, so their scaling dimensions satisfy the bounds
\beq
D(y_1) &=& 1 + \gamma_{y_1}(g=0) \ge  1  \\
D((x_3 z_1)) &=& 1 + \gamma_{(x_3 z_1)}(g=0) \ge 1
\label{nonsat} \\
D(q_3 x_4) &=& 2+ \gamma_{q_3}(g=0)+\gamma_{x_4}(g=0) \ge  1
\label{bound} \\
D(x_4 z_2) &=& 2+ \gamma_{x_4}(g=0) + \gamma_{z_2}(g=0) \ge  1
\label{xzbound}
\eeq
(where $\gamma_\phi$ is the anomalous dimension of the field $\phi$)
with equality holding if the operators are free.  Thus for $g=0$ the first two
interaction terms in the superpotential (\ref{finalspot}) are the products of
three gauge invariant operators, and are thus irrelevant
operators\footnote{They can only be relevant if all three operators are
dimension 1, in which case the operators are free, a contradiction.}.  In other
words, if the coefficients of the first two terms are labeled $\lambda_1/\mu_0$
and
$\lambda_2$ then for $g=0$, $\lambda_1$ and $\lambda_2$ run to zero in the
infrared. Since the fields $(\bar{q} q)$, $(yy)$, and $y_1$
only interact through these irrelevant terms, they are free fields and their
anomalous dimensions vanish.
The effective theory containing the remaining two operators and the $\SU{2}_1$
gauge interactions is in an interacting
non-Abelian Coulomb phase (i.e. it has a non-trivial infrared fixed point at
the origin
of moduli space).
This can be seen by
noting that this is a special case of an \SU{2} theory with $N_F$ flavors (here
$N_F=4$) and trilinear superpotential terms which is dual to an \Sp{2N_F-6}
theory with $N_F$ flavors and trilinear superpotential terms.  The \SU{2}
theory is asymptotically free for $N_F<6$, while the \Sp{2N_F-6}  theory is
asymptotically free for $N_F>18/5$, so  (assuming a la SUSY QCD
that there is a conformal range of fixed point theories between the two
Banks-Zaks \cite{BanksZaks} fixed points\footnote{If the superpotential
couplings
are taken to be arbitrarily small, a Banks-Zaks fixed point can be established
in
perturbation theory at the point where asymptotic freedom is almost lost, then
by
holomorphy \cite{holo} there is a fixed point for arbitrary superpotential
couplings.})
the theory is at an infrared fixed point for $18/5 <  N_F < 6$.  Thus for $g=0$
the
bounds in (\ref{nonsat})-(\ref{xzbound}) are not saturated.  Recall that the
case of five
flavors in the original theory was special because it led to a simple dual
without tensor gauge representations, and it is the absence of tensor
representations that allows for a simple demonstration of an infrared
fixed point for $g=0$.

A few more relations between anomalous dimensions are required to reach some
definite conclusions for non-zero $g$.
Recall that the exact $\beta$ function for the $\SU{2}_1$ coupling is
\cite{NVZ}:
\beq
\beta(g_1) &=& - {{g_1^3}\over{16 \pi^2}} \,\,
{{2 + 5 \gamma_{q_3}(g=0) + 2 \gamma_{x_4}(g=0) + \gamma_{z_2}(g=0) }
\over{1 - {{g_1^2}\over{4 \pi^2}} }} ~,
\eeq
thus at the fixed point:
\beq
0=2 + 5 \gamma_{q_3}(g=0) + 2 \gamma_{x_4}(g=0) + \gamma_{z_2}(g=0) ~.
\label{fixedpoint}
\eeq
Since the last  term in the superpotential (\ref{finalspot}) is a relevant
operator (for $g=0$) with  $R$-charge $2$  the anomalous dimensions satisfy
\beq
\gamma_{(x_3 z_1)}(g=0) + \gamma_{x_4}(g=0) + \gamma_{z_1}(g=0)  = 0 ~,
\eeq
which, with the bound (\ref{xzbound}), implies
\beq
\gamma_{(x_3 z_1)}(g=0) <1~.
\label{compbound}
\eeq
Furthermore $q_3^2$, $x_4^2$, and $q_3 z_1$ are gauge invariant operators
so
\beq
\gamma_{q_3},
\gamma_{x_4} \ge - {{1}\over{2}} ~,
\label{qbound}
\\
\gamma_{q_3} + \gamma_{z_1} \ge -1 ~,
\label{qzbound}
\eeq
independent of $g$.
Combining the fixed point condition (\ref{fixedpoint}) with
the bound (\ref{qzbound}) gives
\beq
2 \gamma_{x_4}(g=0) < -1 - 4 \gamma_{q_3}(g=0)~.
\eeq
This inequality with the bound  (\ref{qbound})  implies
\beq
\gamma_{x_4}(g=0) < {{1}\over{2}}~.
\label{xbound}
\eeq

Returning to the theory with $g(\mu)$ arbitrarily small,
but non-zero, I note that
the anomalous dimensions of $(\bar{q} q)$, $(yy)$, and $y_1$ as well as
the couplings $\lambda_1$ and $\lambda_2$ vanish at $g=0$.  I proceed
by making
the plausible assumption that the anomalous dimensions and
$\beta$ functions of the theory with $g(\mu)$ arbitrarily small
can be reliably analyzed near the scale $\mu$ with a perturbative
expansion in $g(\mu)$.
Thus I am assuming that the anomalous dimensions of the fields with
$\SU{2}_1$ interactions have
reliable perturbative expansions in $g(\mu)$, although I do not know the value
first  ($g(\mu)$ independent) terms in these expansions since  they are
determined by the dynamics of the pure $\SU{2}_1$ fixed point discussed above.%
\footnote{Alternatively they are determined by the (unknown) superconformal
$R$-charge
\cite{Seib,IntSeib}.}
Now consider the running of $g(\mu)$ which is determined by \cite{NVZ}:
\beq
\beta(g) &=& - {{g^3}\over{16 \pi^2}} \,\,
{{4-N + (2N+1)\gamma_{y_1} + 2 \gamma_{x_4} + \gamma_{(x_3 z_1)} }
\over{1 - {{g^2}\over{4 \pi^2}} }} ~,
\nonumber \\
&=& - {{g^3}\over{16 \pi^2}}
\left(4 - N + 2 \gamma_{x_4}(g=0) + \gamma_{(x_3 z_1)}(g=0) \right)
+ {\cal O}(g^5) ~.
\label{betag}
\eeq
The bounds (\ref{compbound}) and(\ref{xbound}) imply that the $\SU{2}_2$
interactions are infrared free ($\beta(g) > 0$) for $N > 6$.

Turning to the superpotential interactions, the $\beta$ functions of the the
first two terms in the
superpotential (expanded to leading order in $\lambda_1$, $\lambda_2$ and
$g(\mu)$)
are:
\beq
\beta_1 & = & 1 + \gamma_{(\bar{q}q)} + \gamma_{q_3}+\gamma_{x_4}+\gamma_{y_1}
\nonumber \\
&=&1 +\gamma_{q_3}(g=0)+\gamma_{x_4}(g=0) +a\lambda_1^2+b\lambda_2^2- c
g(\mu)^2~,\\
\beta_2 &= &\gamma_{(yy)} +2 \gamma_{y_1} \nonumber \\
&=& d\lambda_1^2+e\lambda_2^2-f g(\mu)^2~,
\label{beta2}
\eeq
where $a...f$ are positive numbers.  For the first two terms in the
superpotential to be relevant these two $\beta$ functions
must vanish.  The discussion above indicates that  $\lambda_1$
and $\lambda_2$
must vanish with $g(\mu)$.
Therefore for sufficiently small $g(\mu)$, $\beta_1$ cannot vanish since the
bound
in equation (\ref{bound}) is not saturated.  These considerations suggest that
it is however possible  for $\beta_2$ to vanish. However
the solution of $\beta_2 = 0$ with
$\lambda_1=0$ is that $\lambda_2 \propto g(\mu)$.  Thus in the
infrared limit $\mu \rightarrow 0$ the three couplings
$\lambda_1$, $\lambda_2$, and  $g$ all run to zero.

The conclusion of this analysis is that for $N>6$ the chiral operators
$\overline{q}q  \to   (\overline{q}q)$ and
$\overline{q} A \overline{q}  \to  (yy)$ (as well as the $\SU{2}_2$ gauge
(vector) multiplet and the dual quark $y_1$),
correspond to free fields in the infrared,
while the remaining fields are at an interacting fixed point.  Thus these
theories provide
explicit examples of the phenomena suggested in refs. \cite{LST,selfdual} of a
gauge theory splitting into a free sector and an interacting fixed point sector
in the infrared.
A similar analysis can be applied to the odd case, $N_c=2M-1$, which
can also be shown to have an interacting infrared fixed point for $M > 6$.
Although I have only
proven that the theory with $F=5$ flavors has an interacting infrared fixed
point, I expect that the fixed point will persist up to the point where
asymptotic freedom is lost: $F=2N_c+3$. It should be noted that this analysis
does
not preclude a fixed point for $N \le 6$; to obtain information about these
theories
would require more information about the anomalous dimensions
$\gamma_{x_4}(g=0)$ and $\gamma_{(x_3 z_1)}(g=0)$.  It is suggestive however
that
for $N=1$ and $M=2$ the original theory reduces to vector-like \SU{2} and
\SU{3} theories both of which do indeed have a
non-trivial infrared fixed points.

% ----------------------------------------------------------------------------
\section{Conclusions}
% ----------------------------------------------------------------------------
I have displayed a new dual for \SU{2N} with an antisymmetric tensor, five
flavors, and no superpotential.  Using holomorphy to adjust the ratio of the
scales
of the two gauge groups  in the dual description I have been able to show that
in the five flavor case two
composite ``mesons" become free fields in the infrared, while other degrees of
freedom are
at an interacting infrared fixed point for $N> 6$.  Thus in going from five to
four
flavors (for sufficiently large $N$) the theory goes from a fixed point to
confinement\footnote{Without chiral symmetry breaking.} without passing through
an infrared
free phase.  Such behavior was seen previously \cite{Seib,IntSeib}
in the isolated case of
vector-like \SU{2}, whereas in the generic case of vector-like \SU{N} theories
there is confinement for $F=N+1$ flavors and an infrared free gauge description
for $N+1 < F <3N/2$. (The two bounds coalesce for $N=2$.)
The transition from a fixed point phase directly to a confining phase
as the number of flavors is reduced has been argued to occur\footnote{However
non-SUSY QCD has confinement {\em with} chiral symmetry breaking.} in
(non-SUSY) QCD \cite{QCDFP}.
Given that there is currently no non-perturbative
understanding of non-SUSY \SU{N} gauge theories with an arbitrary number of
flavors, it is somewhat reassuring to find that the expected
behavior of the confinement transition
is actually realized in a large class of theories that {\em are} under
non-perturbative control. However, there is no evidence to suggest that in QCD
there are free, massless composites on the fixed point side of the transition.
On
the contrary the scalar and pseudoscalar (pion) mesonic states are expected
to be massive (and broad) resonances on the fixed point side of the transition
\cite{QCDFP}.  Thus
while some of the qualitative behavior of the confinement transitions in QCD
and
the chiral SUSY theories discussed here is similar, the detailed physics
of the two confinement transitions appears to be quite different.

% ----------------------------------------------------------------------------
\section*{Acknowledgements}
% ----------------------------------------------------------------------------
I would like to thank  Csaba Cs\'aki and Yaron Oz for enlightening discussions
as well as
Martin Schmaltz for enlightening discussions and a critical reading of the
manuscript.
This work was supported by  the National Science
Foundation under grant PHY-95-14797, and also partially supported by
the Department of Energy under contract DE-AC03-76SF00098.

\end{document}